\documentclass[german,a4paper,fleqn,titlepage]{article}
\usepackage{varioref,multicol,epsfig}

\newfont{\myfont}{cmr12 scaled \magstep2}
\newfont{\myfontz}{cmr10 scaled \magstep2}
\newcommand{\figps}[9]    
{
\begin{figure}[htb]
   \begin{center}
      \begin{picture}(#1,#2)(#3,#4)

	\epsfig{file=#9,width=#1cm,height=#2cm}

      \end{picture} \\             %
      \parbox[t]{#5cm}
       {\vspace{-0.5cm}\caption[#8]{\label{#6} #7 \normalsize}}
   \end{center}
\end{figure}
}
\begin{document}
\title{An Electron Bunch Compressor Based on an FEL Interaction in the 
Far Infra Red}
\author{Andreas Gaupp, BESSY}

\date{March 7, 2001}
\maketitle

\begin{large}
\section{Introduction}

The scientific case for a short wave length (photon energy 1~keV and 
more or wavelength 1~nm and less) free electron laser based on the 
principle of stimulated emission of spontaneous emission (SASE FEL) 
is significantly enhanced if the 
optical puls length allowes to resolve molecular vibrations.
The fastest oscillation of interest is the C-H stretching mode at a 
wavelength of $3~\mu$ corresponding to 10~fsec. 
Larger molecular masses oscillate at lower frequencies.
Different proposals exist (at DESY e.g. \cite{Brefeld_1}, \cite{Brefeld_2})
to generate pulses of 20~fsec and less.
In both cases the time structure is derived from ``conventional'' laser 
systems while the electron puls proper is significantly longer.

For an linac based SASE FEL to work electron bunches from the
electron source must be compressed. Conventionally this is done by 
applying an energy chirp along the length of the electron bunch
and subsequent dispersion with momentum compaction.
The bunch length is a function of the applied longitudinal gradient and 
the inital energy spread of the bunch. A typical bunch length is $50~\mu$.

In a free electron laser the interaction of the optical electric field 
with the electron beam is mediated by the undulator magnetic field.
This leads to a (slow) longitudinal motion. Under appropriate 
approximations this motion resembles the well know synchrotron motion
in the high frequency buckets of synchrotrons or linacs. 
The combined action of magnetic and radiation field generates an 
``optical trap'' having longitudinal extend of one wavelength 
$ \lambda _1$. 
The particles execute longitudinal motion 
governed by an equation which is formally identical to the 
well known pendulum equation (see eq. \ref{eq:pendulum} below).
Saturation of the FEL interaction corresponds to half a synchrotron 
period in the optical trap.

In this note an electron bunch compressor is proposed based on FEL type
interaction of the electron bunch with far infrared (FIR) radiation.
This mechanism maintains phase space 
density and thus requires a high quality electron beam to produce 
bunches of the length of a few ten $\mu $.

\section{Theory}

\subsection{Classical FEL Theory}
W. Colson \cite{Colson} summarises the classical one-dimensional FEL 
theory in the limit of the slowly varying amplitude and 
phase approximation. 
The electron dynamics is governed by the pendulum equation
\begin{equation}\label{eq:pendulum}
\stackrel{\circ \circ}{\zeta} = |a| \cos  (\zeta+\phi)
\end{equation}
and the evolution of the radiation field is given by
\begin{equation}\label{eq:field}
\stackrel{\circ}{a}~=~-j<\exp (-i \zeta) >
\end{equation}

$ < ...> $ is averaging over the electron distribution.

The longitudinal coordinate $ \Delta z $ of the electron with respect to the 
synchronous phase is related to the phase $ \zeta $ via the radiation 
wavelength $ \lambda _1 $ by
\begin{equation}\label{eq:zeta} 
\zeta~=~2 \, \pi\, \frac{\Delta z}{\lambda _1} 
\end{equation}
The resonance condition is 
\begin{equation}\label{eq:resonance}
\lambda_1 \ = \ \frac{\lambda _o}{2 \gamma^2}\ (1+K^2)
\end{equation}
The fractional energy deviation  $ \Delta \gamma / \gamma $ from 
resonance is related to the normalised velocity $ \nu $ by
\begin{equation}\label{eq:normalised velocity}
\nu \ = \ 
\stackrel{\circ}{\zeta} \  = \ 4 \pi \, N \ \frac{\Delta \gamma}{\gamma}
\end{equation}
Here $ \stackrel{\circ}{\zeta} $ denotes the derivative of $ \zeta $
with respect to  the normalised time $ \tau $ (see table 1).
The electric field of the radiation is $ E \exp i \phi $.
The dimensionless field amplitude is given by
\begin{equation}\label{eq:field definition}
a~=~(4 \pi N)^2 \ \frac{K}{1+K^2} \ K_1
\end{equation}
with
\begin{equation}
K_1~=~\frac{e\, \lambda_1 \, E}{2 \pi m c^2}
\end{equation}

The dimensionless current density is 
\begin{eqnarray}\label{eq:current density}
j&~=~&32 \ \frac{e^2 \pi ^2}{mc^2} \ \frac{N_e}{\lambda _1}\  
\frac{N^2}{\gamma} \ \frac{K^2}{1+K^2} \\
&~=~&1.85 \cdot \,10^{-2}\ ( \frac{I}{Amp} ) 
\ \frac{N^2}{\gamma}\  \frac{K^2}{1+K^2}
\end{eqnarray}

The Rayleigh range $ z _o $ of an Gaussian mode is related to its waist
size $ w_o $ by
\begin{equation}\label{eq:rayleigh}
w_o \ = \ \sqrt{\lambda _1 \ z _o \, / \,\pi}
\end{equation}

The meaning of most of the other symbols is given in the table.

Helical polarisation is assumed.

\begin{table}[htb]\caption{Symbols} 
\begin{large}
\begin{center}\begin{tabular}{| l | l | l | }
\hline 
e & electric charge & $ 4.8 \cdot 10^{10}  esu $ \\
m & electron mass &   $ 0.911 \cdot 10^{-27} g $ \\ 
c & speed of light & $ 3 \cdot 10^{10}\ cm/sec $ \\
$ \lambda _1 $ & radiation wavelength & \\
$ N $ & number of undulator periods & \\
$ K $ & wiggler parameter & $ \frac{e \lambda _o B}{2 \pi m c^2} $ \\
$ \lambda _o $ & undulator period length & \\
$ B _o $ & undulator field amplitude & Gauss \\
$ K _1 $ & radiation field parameter 
                         & $ \frac{e \lambda _o |E|}{2 \pi m c^2} $ \\
$ E $ & electric field amplitude of radiation & statVolt/cm \\
$N_e $ & number of electrons within one $ \lambda _1 $  & \\
$ \gamma $ & electron energy & \\
$ \tau $ & dimensionless time 0 ... 1 & $(t\, c\, \beta_{||})/(N\, \lambda _o) $
\\
$ \stackrel{\circ}{(...)} $ & derivative with respect $ \tau $ & 
$ \frac{\partial}{\partial \tau} $ \\

\hline
\end{tabular}\end{center}
\end{large}
\end{table}

Particles move in longitudinal phase space $ (\zeta \, , \, \nu) $ 
along trajectories with a constant of motion
\begin{equation}\label{eq:constant of motion}
2 H _o\ = \ \nu _o^2\,- \,2 |a| sin(\zeta _o\, +\, \phi)
\end{equation}

Particles with a constant of motion $  H_o \ \le \ \,a $ stay inside 
the optical trap defined by 
\begin{equation} \label{eq:separatrix}
H_o \ =  \ a
\end{equation}

Equation \ref{eq:separatrix} defines the separatrix with longitudinal 
extend of $ 2 \pi $ corresponding to one wavelength, and vertical 
extend of $ \pm 2 \sqrt{a} $.
Particles near the fixed point at the center of the trap 
at $ (\zeta ,\, \nu ) \ = \ ( \pi /2 ,\ 0) $ execute harmonic oscillations.

\subsection{The FEL Buncher}

The principle of the FEL buncher is to exploit the synchrotron motion in 
the optical trap of a suitably designed FEL interaction to rotate the 
electron bunch in longitudinal phase space by $90^o$.

In the FEL buncher particles are made to execute a quarter of a synchrotron 
period in the optical trap due to an FEL interaction with a superimposed 
radiation field with wavelength $ \lambda _1 $. 
A particle distribution with initial bunch length $\sigma _z$ and 
initial fractional energy spread $ \Delta \gamma / \gamma $
is transfered into a particle distribution of
final length $\sigma _z '$ and final 
energy spread $ \Delta \gamma' / \gamma $.
Provided the initial energy distribution is
small enough and the initial bunch length is short compared to the 
extend of the trap this causes bunching of the beam 
(see fig. \ref{fig:phasespace_1}).

We assume that all motions in the optical trap are harmonic, i.e. the 
inital phase space distribution of the beam occupies only a small 
fraction of the optical trap near its center.
The equation of motion of the field amplitude eq. \ref{eq:field}
is not considered here assuming no change of the electric field. 

\figps{14}{14}{0}{0}{10}{fig:phasespace_1}
{Longitudinal phase space $\nu$ versus $\zeta$ of FEL interaction with initial (open symbols)
and final (full symbols) particle distribution.
The initial bunch length is $ \sigma_z~=~0.16 \, \lambda _1$,
the initial energy spread is 
$ \frac{\Delta \gamma}{\gamma}~=~5 \cdot 10 ^{-4} $,
the number of periods is N~=~64,
the normalised field strength is $a~=~1.35 \cdot (\pi/2)^2 $.
}{phasespace}
{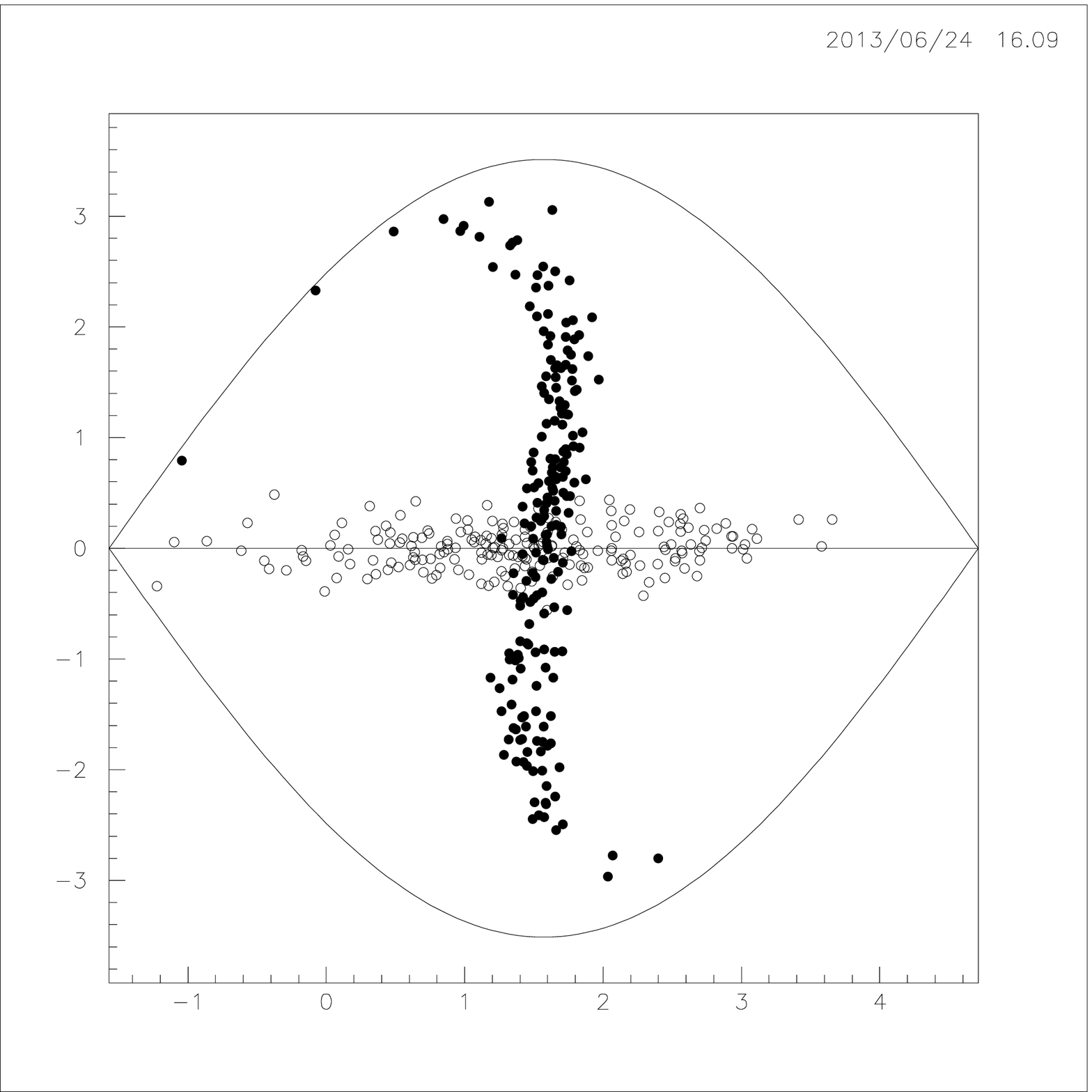}

Conservation of phase space density implies
\begin{equation}\label{eq:phase space density}
\frac{\Delta \gamma '}{\Delta \gamma} ~= ~ \frac{\sigma _z}{\sigma _z '}
\end{equation}

The bunching condition requires a quarter of a synchrotron period in 
the optical trap.
From eq. \ref{eq:pendulum} we note that the oscillation frequency for small 
amplitudes is $ \sqrt{a} $ and the phase advance is $ \sqrt{a}\, \tau $.
At the end of the interaction $ \tau\, = \, 1 $.
The {\bf bunching condition } thus is
\begin{equation}\label{eq:bunching condition}
\sqrt{a}~=~\pi /2
\end{equation}

Note: Eq. \ref{eq:bunching condition} refers to the final synchrotron 
phase advance. 
A variation of the field amplitude during the interaction
can be tolerated.

There are two further requirements:

The electron bunch must be short compared to the wavelength. 
This is needed to achieve the desired bunch length.
Otherwise non-linear motion far off the center of the trap dilutes the 
phase space distribution and reduces the bunching effect. 
The {\bf short bunch requirement} reads
\begin{equation}\label{eq:short bunch}
\frac{2 \pi \,\sigma _z }{\lambda _1 }\ \leq \ 1
\end{equation}
The induced energy spread is
\begin{equation}\label{eq:energy spread}
\Delta \gamma ' / \gamma ~  \geq \frac{1}{4 \pi N} 
\ \frac{\sigma _z } {\lambda _1 / 2 \pi}
~=~\frac{1}{2N}\ \frac{\sigma _z}{\lambda _1}
\end{equation}

The {\bf cold beam requirement} is needed to get the desired puls length
\begin{equation}\label{eq:cold beam}
\sigma _z '~ \geq 4 \pi N \ \frac{\Delta \gamma}{\gamma}  \ 
\frac{\lambda _1}{2\pi} 
\ = \ 2\,N\,\frac{\Delta  \gamma}{\gamma} \, \lambda _1
\end{equation}
or
\begin{equation}\label{eq:N}
N\ \leq \ \frac{1}{2} \ \frac{\sigma _z '}{\lambda _1} \ 
\frac{ \gamma}{ \Delta \gamma}
\end{equation}

\subsection{Power Requirement and Jitter}

A major limitation of the FEL buncher is the required optical power to 
drive the FEL interaction. 

The FEL interaction strength depends on the focussing of the optical 
mode. The interaction strength is maximised for a Gaussian mode with
a Rayleigh range $ z_o $ of
\begin{equation}\label{eq:Rayleigh range}
z_o~ = \frac{1}{2} N  \lambda _o
\end{equation}
The minimum optical cross section is
\begin{equation}\label{eq:waist}
 w_o~=~ \sqrt{z_o \lambda _1/\pi} 
\end{equation}
The cross section varies along the axis z as 
\begin{equation}\label{cross section}
w^2(z)~=~w_o^2\,(1\,+\,(\frac{z}{z_o})^2)
\end{equation}

The power in the optical driving field is using eq. \ref{eq:Rayleigh range},
\ref{eq:waist} and \ref{eq:resonance}
\begin{eqnarray}\label{eq:power}
 P &~=~& \frac{c}{2}\, E^2 \  \cdot \pi\, w_o^2  \\
&~=~& (\frac{1}{4 \pi})^4 \ \frac{c}{2}\ (\frac{2 \pi m c^2}{e})^2 \ 
\frac{1+K^2}{K^2} \ \frac{\gamma ^2}{N^3} \ a^2
\end{eqnarray}

The field strength $ E $ follows from the bunching condition eq. 
\ref{eq:bunching condition}.
We obtain
\begin{eqnarray}
P&~=~&\frac{c}{2^{13}}\ (\frac{2 \pi m c^2}{e})^2 \ \frac{\gamma^2}{N^3}
\ \frac{1+K^2}{K^2}  \\
&~=~& 42.2\,MW~ \frac{\gamma^2}{N^3} \  \frac{1+K^2}{K^2}
\end{eqnarray}

Using eq. \ref{eq:N} the power becomes
\begin{eqnarray}\label{eq:power numerical}
P&~=~& \frac{c}{2^{10}}\ (\frac{2 \pi m c^2}{e})^2
\ ( \frac{\lambda _1}{\sigma _z'} )^3
\ \frac{ \Delta \gamma ^3}{\gamma} \ \frac{1+K^2}{K^2} \\
&~=~& \ 340\,MW \ ( \frac{\lambda _1}{\sigma _z'} )^3
\ \frac{ \Delta \gamma ^3}{\gamma} \ \frac{1+K^2}{K^2}
\end{eqnarray}

Using the equality in eq. \ref{eq:short bunch} shows the scaling of the 
optical power with the third power of the compression ratio times the 
energy spread:
\begin{eqnarray}\label{eq:power scaling}
P&~=~& \frac{\pi^3}{2^7} \ c \  (\frac{2 \pi m c^2}{e})^2 
\ ( \frac{\sigma _z}{\sigma _z'} )^3
\ \frac{ \Delta \gamma ^3}{\gamma} \ \frac{1+K^2}{K^2} \\
&~=~& \ 84\,GW\ ( \frac{\sigma _z}{\sigma _z'} )^3
\ \frac{ \Delta \gamma ^3}{\gamma} \ \frac{1+K^2}{K^2} 
\end{eqnarray}
This power must be increased to compensate for non-harmonic particle 
motion in the optical trap.

A higher electron energy linearly decreases the power requirement at the 
expense of a longer interaction region (cf. eq. \ref{eq:resonance}).

The resulting electric field strength of a beam  with cross section $ A $ is
\begin{equation}\label{field strength}
E~=~\sqrt{\frac{2\,P}{\epsilon _o\,c \, A}}
\end{equation}
($\epsilon _o~=~8.85 \cdot 10^{-12}\ F/m $)

At the waist we have $ A~=~\pi \, w_o^2~=~z_o \, \lambda _1 $.

The electron bunch slips back by one optical wavelength per undulator 
period. 
The optical puls  must have a length of at least this slippage distance
$ N \cdot \lambda _1 $. The total optical energy in the puls thus is
\begin{equation} \label{eq:energy}
E _{opt}~=~\frac{1}{c} \, P\,N\, \lambda _1
\end{equation}

A further limitation is due to the short bunch requirement 
eq. \ref{eq:short bunch}. It refers to an effective bunch length 
including jitter between the phase $ \phi $ of the optical field and the 
electron bunch position $ \zeta _o $. For reliable operation one must 
have $ jitter \ll \lambda _1 $. This limits the usable wavelength and 
thus the achievable effective gradient.

\section{A Numerical Example}

Designing an FEL buncher at this stage consists of the following steps:
\begin{enumerate}
\item Select a wavelength $ \lambda _1 $ such that the short bunch requirement 
is met.
\item Select a position along the linear accelerator 
and get the power from eq. \ref{eq:power numerical}.
The short bunch must be transportable to its final destination 
without unacceptable bunch lengthening.
If needed the electron bunch can be cooled by scraping off particles with a 
large energy deviation prior to the FEL buncher.
\item Check that the resulting energy spread eq. \ref{eq:energy spread}
is acceptable.
\item Check that the number of periods from eq. \ref{eq:N}
is acceptable.
\item Check the technical feaseability of the (helical) undulator.
\item Design a suitable source for the optical field.
\end{enumerate}

As an example we choose the situation downstream of the third bunch 
compressor of the  TTF-FEL as described in the proposal \cite{TTF-FEL}.

The wavelength is chosen to be 
$ \lambda _1~=~314\, \mu$, which is $ 2 \pi $ the initial bunch length.
This choice is dictated by the short bunch requirement and the desire to 
avoid unharmonic motion of the bunch in the optical trap.
The jitter must be well below 1~psec.
This is believed to be achievable if the optical
phase is synchronized to the main electron beam via the FIR generator
(cf. below).

The electron beam energy is 516 MeV ($\gamma~=~1010$)
with an initial energy spread of 1000~keV. 
In the interest to limit the power we scrap off particles such that the
energy spread is reduced to 
$ \Delta \gamma/ \gamma~=~ 5 \cdot 10^{-4} $ 
and $ \Delta \gamma\,=\,0.5 $
($ \sigma _{\nu}~=~ 4 \pi \,N\, \frac{\Delta \gamma}{\gamma}~=~0.20 $).
This reduces the bunch charge by a factor of about 4.
-- The initial bunch length is $ \sigma_z~=~50~\mu$.

Aiming at  final bunch length of $ \sigma _z'~=~20~\mu $ 
corresponding to $60~fs$ 
the power must be $ P~=~300\,MW$.
This number includes a factor of 1.8 to compensate for the non-harmonic 
motion of particles in the optical trap.

The number of periods  from eq. \ref{eq:N} is $N~=~64 $.
The energy in the optical puls of length $N \cdot \lambda _1 $ is
$ E_{opt}~=~0.020~J $.
The final energy spread is 
$ \frac{\Delta \gamma}{\gamma}~=~1.2 \cdot 10^{-3} $.

The helical undulator has a period of $\lambda_o~=~400~mm$ and a 
field amplitude of $B_o~=~10.6~kGauss$ resulting in a wiggler strength 
$K~=~40$. 
The interaction length is thus about 26~m.
There is a copious amount of spontaneous hard radiation.

The Rayleigh range of the optical $TEM_{oo} $ mode is $z_o~=~13\, m $
leading to a waist of $ w _o~=~36\, mm $.

The electric field strength at the waist E~=~6.1 MV/m.


Fig. \ref{fig:phasespace_1} shows a one-dimensional simulation (constant 
field amplitude through out the interaction) integrating eq. 
\ref{eq:pendulum} for 200 particles with the parameters given above.

The discussion of a possible generator is deferred to the next section.

The requirement will be relaxed if the energy spread and/or bunch length
of the initial electron beam is further reduced.
The interaction length is reduced for an earlier location in the 
linac.

\section{The Generator}

The required power of the order of GigaWatt increases rapidly if the energy 
spread is bigger than assumed and/or if stronger bunching is required.
These power levels suggest to use a FEL type generator.
At these FIR wavelength an electron beam is easily
pre-bunched, i.e. the bunch length is much shorter than the wavelength.

\subsection{Secondary Electron Beam}

We first consider a generator operating on a secondary electron beam
which can be discarded once the power is generated.

Both beams, the secondary one and the main one, consist of a single 
electron bunch. The secondary bunch is synchronized to the main bunch.
The secondary bunch comes from a laser driven photo cathode rf gun 
much like the main beam. The two guns a synchronized by deriving the laser 
pulses from a common laser oscillator system.

The undulator of the generator is of the same length as the buncher 
undulator. 
Thus the emitted radiation puls has the proper length for the FEL 
buncher.

The spontaneous radiation generated in this undulator 
is much stronger than the usual undulator radiation since the electron 
is already pre-bunched.
The spontaneous radiation is amplified in further undulator modules 
of the same length. 
Phasing, dispersion, and focussing of the radiation is achieved 
using retarders in between the modules.

The electron beam should be defocussed in the undulators to efficiently couple 
to the lowest optical mode ($ TEM_{oo} $), and should be focussed to stay 
clear from the components of the retarder.

There are at least two options for the retarder.

The first one consists of plates of appropriate thickness and material 
to slow down the radiation.
To achieve a retardation by $ \Delta~=~N \cdot \lambda_1~=~20~mm $ 
the plate should have  a thickness of $ \Delta/n $ where
$ n $ is the index of refraction.
These plates have curved surfaces introducing focussing to counteract 
diffraction. They contain a hole in the center to transmit the (focused) 
electron beam.
Further units are needed to controll phasing , i.e. a phase modulator 
acting on the electron beam much like in an optical klystron.
The undulator modules are tapered to match energy extraction.

A major problem with this design is the power density 
in the focussing retarders.

The second option for the retarder consists in a 
zig-zag arrangement of the undulators.
The electron beam is bent by magnets while the light beam is reflected by 
a focussing mirror. 
The straight path of the light is longer than the curved path of the 
electrons. This path lengthening must be equal to
$ N \cdot  \lambda _1$.
Steerer magnets associated to the bending magnets can vary the electron
beam path length.

The induced path length difference $ \Delta $ at a bending radius R 
and a deflection angle $ \theta $ is 
\begin{equation}\label{eq:path length difference}
\Delta~=~ 2\, R \ (\tan{\theta/2}\,-\, \theta/2)
\end{equation}
At $R~=~1.6 \,m $ a deflection angle of $ \theta~=~30^o $ leads to a path 
difference of $ \Delta~=~N \cdot \lambda _1~=~20~mm$.

The separation $ x $ between the optical beam and the electron path is
\begin{equation}\label{eq:stay clear}
x~=~ R \, (\,\frac{1}{\cos{\theta/2}}\,-\,1)
\end{equation}
The maximum seperation between the optical beam and the electron beam
becomes $x~=~56~mm$.

\subsection{Main Electron Beam FEL}

In this case the main electron beam is used to generate the required FIR 
power in a FEL cavity. 
The electron beam consists of a sequence of bunches. 
The first few bunches of the bunch train are used to build up the 
intracavity power. The trailing bunches are compressed in the FEL 
interaction and simultaneously provide for enough optical gain to maintain 
the power level corresponding to optimum bunching. 

To do this the beam has to have a repetition rate matched to the cavity 
length. The choice of the cavity length is probably dictated by the 
consideration of the power density on the mirror surfaces (see below).

This FEL uses a prebunched electron beam.
\figps{14}{14}{0}{0}{10}{fig:phasespace_2}
{Longitudinal phase space $\nu$ versus $\zeta$ of FEL interaction with initial (open symbols)
and final (full symbols)  particle distribution. 
Initial distribution off center.
The initial bunch length is $ \sigma _z~=~0.125 \cdot \lambda _1$,
the initial energy spread is 
$ \frac{\Delta \gamma}{\gamma}~=~1 \cdot 10^{-3} $,
the number of periods is N~=~16, and the normalised field strength is 
$ a~=~1.25 \cdot (\pi/2)^2 $.
}
{phasespace}{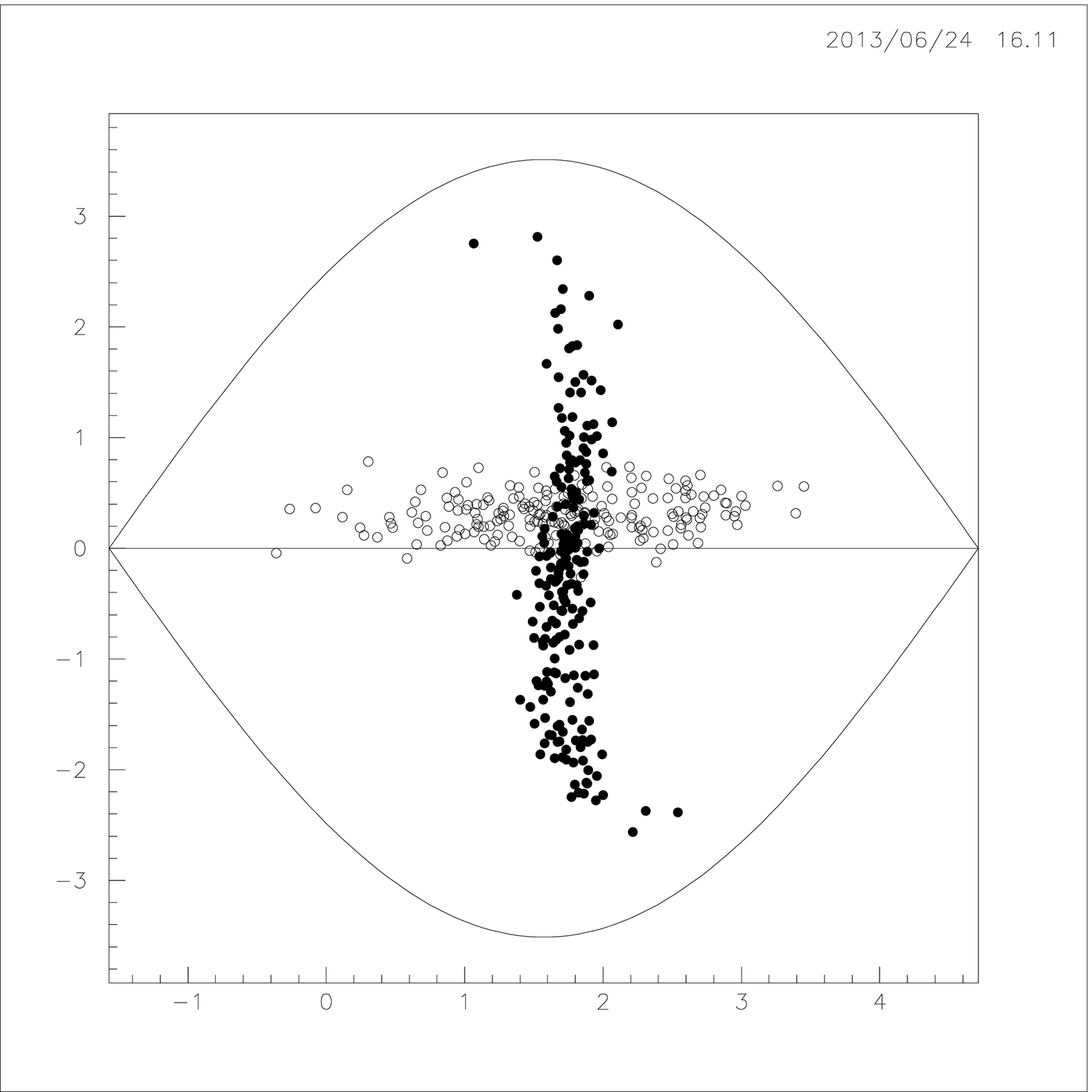}
%
The losses in the cavity and thus the intra-cavity power density 
is regulated such that in the steady state 
the electrons perform a quarter of a synchrotron oscillation 
inside the optical trap.
It is speculated that the
phase of the electron bunch can be stabilised by choice of cavity length 
and magnetic field amplitude
as  qualitatively shown in fig. \ref{fig:phasespace_2}.
Also optical dispersion in the cavity can help.
It needs to be verified by (numerical) 
simulations including the nonlinear electron dynamics.

We note that the coupling between electron beam and the optical mode is 
rather poor. The filling factor i.e. the ratio of the cross sections of 
the electron beam and the optical mode is much less than 1. 
In other words the 
spatial disptribution of the generated radiation does not match the 
field distribution in the cavity very well.

Fig. \ref{fig:phasespace_2} shows a one dimensional simulation with the 
following parameters:
The wavelength is $ \lambda _1 ~=~400\ \mu$;
the undulator has $N~=~16$ periods with a period length of
$ \lambda _o~=~500\ mm$;
the wiggler strength is $K~=~40$ corresponding to a field amplitude of 
$ B_o~=~0.86\ Tesla$;
the Rayleigh range is $z_o~=~4\ m$ and the waist size is 
$ w_o~=~22.5 \ mm$.
The electron beam has an energy of 516~MeV ($ \gamma~=~1010 $)
and is scraped to an energy spread of $ \Delta \gamma / \gamma~=~
1 \cdot 10^{-3}$.
A bunch charge of $0.5~nCb$ and a bunch length of $ \sigma_z~=~0.125 \cdot 
\lambda_1~=~50~\mu m$ represents a current of $3~kAmp$. This translates 
into a dimensionless current density of $j~=~14$; thus the neglect of 
the field driving term in equ. \ref{eq:field definition} is a crude 
approximation.
The intracavity circulating power is $P~=~16\ GW $. 
The power corresponds to $a~=~1.25 \cdot (\pi/2)^2 $.
This increase beyond equ. \ref{eq:bunching condition} 
compensates for the non-harmonic particle motion in the optical trap.
The electric field strength at the waist reaches $E~=~88\ MV/m  $; 
the optical puls energy is $E_{opt}~=~350\ mJ $. 
 
The parameters leading to the situation in fig. \ref{fig:phasespace_2}
cause a change in the electron energy of
$ \frac{\Delta \gamma}{\gamma} ~=~ \frac{\Delta \nu }{4 \pi \, N}
~=~2.8 \cdot 10^{-3} $. 
At a bunch charge of $ 0.5~nCb $ the optical gain at an optical power 
level of 16~GW is
$ g ~=~ \frac{electron~energy~change}{optical~energy}
~=~ 2 \cdot 10^{-3} $.
It thus takes about 500 electron bunches to increase the power by a 
factor $e $ (assuming no losses) at the desired steady state.
It is speculated here that the FIR power in the cavity has built up 
from the spontaneous emission to the desired power level before 
the end of the linac bunch train.

The observed bunch length is $ \sigma _z '~=~10\,\mu $ 
in this simulation and the final energy spread is 
$ \Delta \gamma ' / \gamma~=~4 \cdot 10^{-3}$.

At the time of the writing the principal limitation seems to be the 
power density on the cavity mirrors which can only be reduced by 
designing a long cavity.
When this problem can be solved the bunching can be enhanced by using a 
shorter interaction length and a higher power level 
which at the same time tolerates a larger 
initial energy spread.

\section {Summary and Outlook}

The proposed FEL buncher uses the transvers electric field of a freely
propagating radiation field and couples it to the electron beam
via the FEL interaction.
Since the wavelength can be much shorter than in an RF device the 
potential for a high longitudinal gradient is given. 

However, the reqirement for the optical power density inside the
FEL interaction is very significant.
and a more realistic study than the one done here will further
increase the power requirement.
The calculations presented here are based on the pendulum equation 
eq. \ref{eq:pendulum} which depends on the slowly varying amplitude 
and phase approximation.
The simulation assumes a constant optical field, i.e. no diffraction. 
It also assumes perfect matching of the electron beam cross section to 
the optical mode size. 
In reality a mismatch of the order of 1000 is probable. 
This does not change the coupling of the electron beam to the optical 
field in the buncher, but it reduces the gain in the generator, 
in particular in the case of fig. \ref{fig:phasespace_2} where 
the main beam is used to generate the FIR power.

Handling the high optical power in the GW range requires special 
attention. It is believed that the power handling capability of mirrors 
is limiting. 
Extending the cavity length does distribute the power over a larger area
but requires an increased mirror size. 
Consider the example of fig. \ref{fig:phasespace_2}:
The power density in the waist is about $ 1~GW/cm^2 $. When a deduction 
of the power density by a factor of 100 is required, the cavity length 
must be 80~m and the mirror diameter becomes 1.4~m.

Spiking is undesirable from the users' point of view.
Spiking in a high power SASE FEL is due to the combination of synchrotron
motion in the optical trap and the lethargy of the FEL interaction.
While seeding improves the spectral line width by correlating the 
optical phase of different spikes it does not in general eliminate spiking
as such.
Short electron bunches may not only be beneficial from the point
of view of doing time resolved experiments 
but also from the point of view of controlling the spiking.
Spiking is expected to be reduced by an electron bunch length
comparable to the total slippage distance.
An electron  bunch length of $ 20~\mu $ may influence spiking up to 
a photon energy of about 100~eV.

Alternatively, the synchrotron motion in the optical trap
is significantly altered by 
tapering the magnetic field along the length of the SASE undulator
thus influencing spiking.

\end{large}
\vfill
\today \\
userdisk\_5:[gaupp.texte]FEL\_Buncher.tex

\vfill


\newpage

\begin{thebibliography}{99}

\bibitem{Brefeld_1}
W.Brefeld, B.Faatz, J.Feldhaus, M.K\"orfer, J.Krzywinski, T.M\"oller, J.
Pflueger, J.Rossbach, E.L.Saldin, E.A.Schneidmiller, M.V.Yurkov,
``Generation High Power Femtosecond Pulses by Subharmonically Seeded 
Soft X-ray FEL at DESY'', DESY 2000, unpublished.
%
\bibitem{Brefeld_2}
W.Brefeld, B.Faatz, J.Feldhaus, M.K\"orfer, J.Krzywinski, T.M\"oller, J.
Pflueger, J.Rossbach, E.L.Saldin, E.A.Schneidmiller, S.Schreiber,
M.V.Yurkov,
``Development of a Femtosecond Soft X-Ray SASE FEL at DESY '', 
 DESY 2000, unpublished.
%
\bibitem {Colson}
W.B.Colson ``Classical Free Electron Laser Theory'', Chapter 3 in ``Free 
Electron Laser Handbook'', ed. Colson, Pellegrine, Renieri, North 
Holland 1988
%
\bibitem{TTF-FEL}
``A VUV Free Electron laser at the TESLA Test Facility at DESY, 
Conceptual Design Report'', DESY Print TESLA-FEL 95-03, June 1995
\end{thebibliography}
\end{document}